\documentclass[aps,prx,letterpaper,superscriptaddress,longbibliography, 12pt]{revtex4-2}

\usepackage{amsmath,amsthm,amsfonts}
\usepackage{multirow}
\usepackage{array}
\usepackage{soul}
\usepackage{float}
\usepackage{graphicx}
\usepackage{afterpage}
\usepackage[cdot,amssymb]{SIunits}

\usepackage[justification=raggedright,singlelinecheck=false,labelfont=bf,labelsep=quad, format=plain,font=small]{subcaption}
\usepackage[colorlinks,citecolor=blue,urlcolor=blue,bookmarks=false,hypertexnames=true]{hyperref}
\usepackage{cleveref}
\usepackage[justification=raggedright,singlelinecheck=false,labelfont=bf,labelsep=quad, format=plain,font=small]{caption,subcaption}
\usepackage{booktabs}

\begin{document}

\title{Atomic layer etching of InGaAs using sequential exposures of atomic hydrogen and oxygen gas}

\author{Mete M. Bayrak}
\affiliation{Division of Chemistry and Chemical Engineering, California Institute of Technology, CA 91125, USA}
\author{Anthony J. Ardizzi}
\affiliation{Division of Engineering and Applied Science, California Institute of Technology, Pasadena, CA 91125, USA}
\author{Sadhvikas Addamane}
\affiliation{Center for Integrated Nanotechnologies, Sandia National Laboratories, Albuquerque, NM 87185, USA.}

\author{Kieran Cleary}
\affiliation{Cahill Radio Astronomy Lab, California Institute of Technology, Pasadena, CA 91125, USA}
\date{\today}

\author{Austin J. Minnich}
\email{aminnich@caltech.edu}
\affiliation{Division of Engineering and Applied Science, California Institute of Technology, Pasadena, CA 91125, USA}
\date{\today}

\begin{abstract}
The high frequency performance and yield of III-V semiconductor devices such as InP HEMTs is negatively impacted by subsurface etch damage and non-uniform etch depth over the wafer.  Atomic layer etching (ALE) has the potential to overcome this challenge because of its ability to etch with Angstrom-scale precision, low damage, and intrinsic wafer-scale uniformity. Here, we report an ALE process for InGaAs based on sequential atomic hydrogen and oxygen gas exposures. An etch rate of 0.095 \AA/cycle was observed at 350 \degreecelsius~using ex-situ spectroscopic ellipsometry. The sample remains atomically smooth after 200 cycles of ALE. This process could be employed as a gate recess etch step in InP HEMT fabrication to improve microwave performance and yield.


\end{abstract}

\maketitle

\newpage

\section{Introduction}


III-V compound semiconductors such as InGaAs and InAlAs are of high importance for both fundamental science and applications. In particular, high electron mobility transistors (HEMTs) based on InGaAs/InAlAs quantum wells are widely used for remote sensing \cite{deal_inp_2012}, radio astronomy \cite{pospieszalski_extremely_2018,chiong_low-noise_2022}, communications \cite{hamada_millimeter-wave_2019,park_sub-50_2023}, and increasingly quantum computing due to their outstanding noise performance \cite{bardin_cryogenic_2021,bardin_microwaves_2021,li_investigation_2024}. 
A critical step in HEMT fabrication is the gate recess etch in which the cap (consisting of highly doped InGaAs) and a portion of the barrier (undoped InAlAs) are etched to place the gate electrode as close as possible to the conducting channel while maintaining a low leakage current ($\lesssim 1$ $\micro$A/mm) \cite{saranovac_pt_2017,li_influence_2022}. The desired etch depth precision is around a nanometer over the wafer, which is challenging to achieve with typical wet or dry etching due to local variations in etch rate. Etching non-uniformities degrade yield, and etch-induced damage limits the degree of scaling that can be achieved, which in turn sets high frequency performance limits \cite{yun_impact_2018}. Etch damage also leads to the formation of electrical traps at the gate-barrier Schottky junction which causes gain instabilities \cite{christensson_low_1968,van_der_ziel_unified_1988,weinreb_multiplicative_2014}. 


Atomic layer etching (ALE) is an emerging nanofabrication technique that consists of a set of sequential and self-limiting half-steps that together produce etching \cite{kanarik_atomic_2018}. ALE is analogous to atomic layer deposition (ALD) except that the half-steps produce etching rather than deposition. An ALE process generally consists of a modification step followed by a selective removal of the modified surface layer. The removal of this modified layer defines whether the ALE is thermal (isotropic) or directional (anisotropic). Directional ALE was demonstrated first, and it exploits the difference in surface binding energy to sputter only the modified layer with ions or neutral atoms \cite{doi:10.1021/acs.jpclett.8b00997}. In 2015, thermal ALE based on ligand-exchange reactions was first reported \cite{lee_atomic_2015-1}.  Since then, several other approaches for thermal ALE have been reported.   


Thermal ALE of InGaAs has previously been demonstrated using vapor HF to form fluoride compounds and ligand-exchange for volatilization using dimethylaluminum chloride (DMAC) \cite{doi:10.1021/acs.nanolett.9b01525}. This process had an etch per cycle (EPC) of around 0.1 \AA/cycle and was used to fabricate an InGaAs FinFET with record performance  \cite{8614536}. Directional ALE of InGaAs has also been explored using chlorine radicals as the modification step and low energy Ar ions to sputter the modified surface layer \cite{Park_2017}. This process was reported to achieve an EPC of 1.1 \AA/Cycle and lower surface roughness than was achieved with reactive ion etching. Despite these advances, processes which preserve the stoichiometry of the ternary alloy while inducing as little surface damage as possible remain of interest. 

Here, we report an ALE process for InGaAs based on sequential exposures of hydrogen radicals and oxygen gas.  The process exhibits an etch rate of around 0.1 \AA/cycle. Both half cycles are observed to be self-limiting. The process preserves the atomic-scale smoothness of the epitaxial wafer over 200 cycles. XPS analysis indicates that In compounds are preferentially etched over Ga compounds. The process has potential for integration into InP HEMT fabrication to enable improved yield and high-frequency microwave performance.

\section{Methods}

The samples consisted of an epitaxial stack of InGaAs and InAlAs films grown on an InP substrate by molecular beam epitaxy. The top film was a 200 nm thick lattice-matched n-In\textsubscript{0.53}Ga\textsubscript{0.47}As layer with carrier concentration of $7.5 \times 10^{18}$ cm\textsuperscript{-3}, below which was a 200 nm lattice matched In\textsubscript{0.52}Al\textsubscript{0.48}As buffer. The films were grown on 2 inch InP wafers and diced into $10 \times 10$ mm chips. Prior to use, the chips were spin cleaned with acetone followed by isopropyl alchohol at 4000 rpm.  

ALE experiments were carried out in an Oxford Instruments FlexAL II atomic layer deposition tool. Hydrogen plasma was generated through an inductively coupled plasma (ICP) source. To limit ion damage to the chip, a pseudo-remote plasma was achieved by encasing the chip in a roof structure to eliminate direct line of sight between the chip and the ICP plasma source. The roof structure was a $30 \times 30$ mm sapphire chip supported by two $10 \times 30$ mm sapphire chips, forming a channel in which the chip sits. This structure allows gas reactants to diffuse to the chip while minimizing direct ion impingement.


Thickness changes due to etching were measured by ex-situ ellipsometry using a J.A. Woolam M-2000 spectroscopic ellipsometer. Measurements were taken in the center of the chips in a line perpendicular to the channel of the roof structure with points spaced 1 mm apart to avoid edge effects from the roof structure. Chips were placed in the same position and orientation before and after etching. Ellipsometry data was taken between 370-1000 nm at angles of 60, 65, and 70 degrees. The data was fit using J.A. Woollam CompleteEASE  software with a model consisting of InGaAs oxide, In\textsubscript{0.53}Ga\textsubscript{0.47}As, In\textsubscript{0.52}Al\textsubscript{0.48}As, and InP.

Before ALE, preconditioning of the chamber was performed to ensure a consistent initial chamber condition. The vacuum chamber chuck was heated to the process temperature of $350$ \degreecelsius~to allow it to thermalize while the rest of the preconditioning was carried out. Next, 6 minutes of Ar/O\textsubscript{2}/SF\textsubscript{6} plasma (120 sccm Ar, 40 sccm SF\textsubscript{6}, 40 sccm, O\textsubscript{2}) was performed, followed by 300 cycles of alumina ALD ($\sim 35$ nm). Alumina was chosen as it is does not react with the oxygen gas and hydrogen plasma used during ALE. Finally, 50 cycles of the ALE recipe was performed, after which the chip was loaded into the reactor.  

\begin{figure}
    \centering
    {\includegraphics[width=\textwidth]{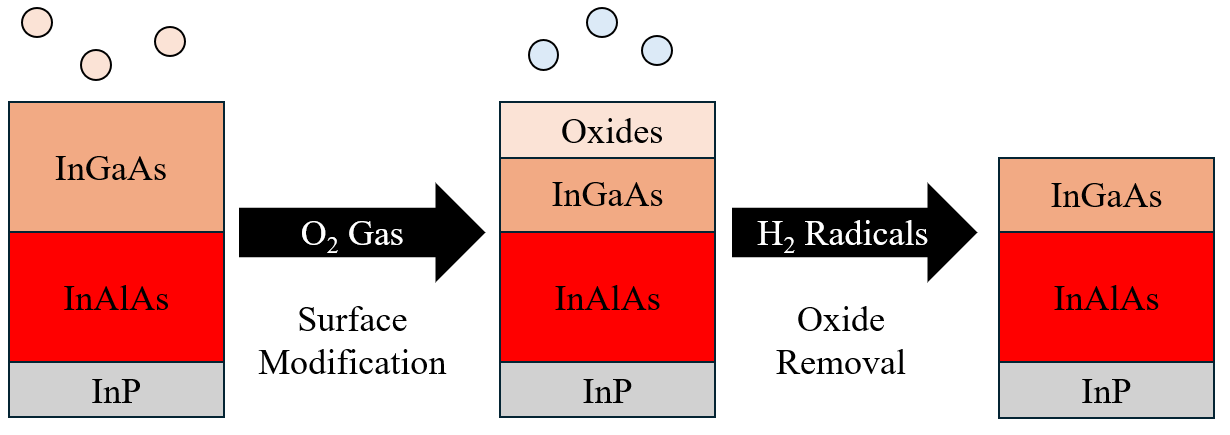}}
    \caption{(a) Schematic of the ALE process. InGaAs is oxidized by oxygen gas (light orange dots). This oxide layer is then removed by hydrogen radicals (blue dots).  
    }
    \label{fig:ALE_recipe}
\end{figure}

The chip was then allowed to heat to $350$ \degreecelsius~for 4 minutes while Ar was flowed at 100 sccm, with a chamber pressure of 100 mTorr. The ALE process was then carried out as shown in \Cref{fig:ALE_recipe}. First, 50 sccm of H\textsubscript{2} was flowed for 10~seconds, reaching a stable chamber pressure of 100~mTorr after which a plasma was struck at 100~W and held for 2 minutes. The chamber was then purged for 2 seconds with 200 sccm of Ar and then pumped down for 10 seconds. This initial 2 min H\textsubscript{2} exposure was used to remove the oxide formed when the sample is removed from the chamber to perform ex-situ ellipsometry. 

The main ALE cycle was then carried out as follows. The oxidation half step was performed by flowing 100 sccm of O\textsubscript{2} for 45 seconds reaching a chamber pressure of 225 mTorr. The chamber was then purged by flowing 300 sccm of Ar for 3 seconds and then pumping down for 10 seconds. After this, the hydrogen plasma half cycle was carried out by first flowing 50 sccm of H\textsubscript{2} for 10 seconds at 100 mTorr and then striking and holding a 100 W plasma for 1 minute. The chamber was then purged with Ar and then pumped down for 10 seconds before restarting the ALE cycle. After processing the chamber was pumped down for 1 minute before the chip was unloaded to the load-lock. The chip was then held in dry nitrogen atmosphere in the load-lock for 5 minutes to cool.


The saturation curves were acquired at constant chamber temperature, pressure, and ICP plasma power, only varying the times of the oxygen gas and hydrogen radical exposures in their respective saturation curves while holding the other constant. The same purging and pumping was used for all processes. The saturation curves were obtained using a single sample. After the chips were preconditioned, the initial film thicknesses were determined via ellipsometry. The chips then underwent 100 cycles of ALE for a given half cycle exposure time before being measured again by ellipsometry. This process was repeated to generate the saturation curve.  

Surface roughness measurements were taken using a Bruker Dimension Icon atomic force microscope (AFM). PeakForce tapping mode was used with a ScanAsyst-Air AFM probe. A $200 \times 200$ nm square region was scanned by segmenting the region into 1024 lines each of which contains 1024 points where data was collected. The raw height maps were processed using the NanoScope Analysis software from Bruker. This raw data was processed using a 2D flatten and low pass filter to remove tilt angle and scan artifacts. The power spectral density (PSD) was then calculated using Nanoscopes's 2D isotropic PSD feature.

Surface characterization was performed via ex-situ X-ray photoelectron spectroscopy (XPS) using a Kratos Axis Ultra XPS with a monochromatic Al K$\alpha$ source. To perform depth profiling, an Ar ion beam was used to mill the sample in 30 s cycles, and data were collected for 8 subsurface points. The XPS data was processed using CASA-XPS from Casa Software Ltd. The backgrounds were taken to be Shirley and the line-shape of the In, Ga, and As peaks were taken to be a mix of Gaussian and Lorentzian. The exact ratio of the line-shape was determined by minimizing the residual of the individual peaks in the bulk where no oxide was present. To compare atomic percentages, the Scofield relative sensitivity factors were used, which recovered the expected In:Ga ratio of 1.13 in the bulk.


\section{Results}

\begin{figure}
    \centering
    {\includegraphics[scale=.65]{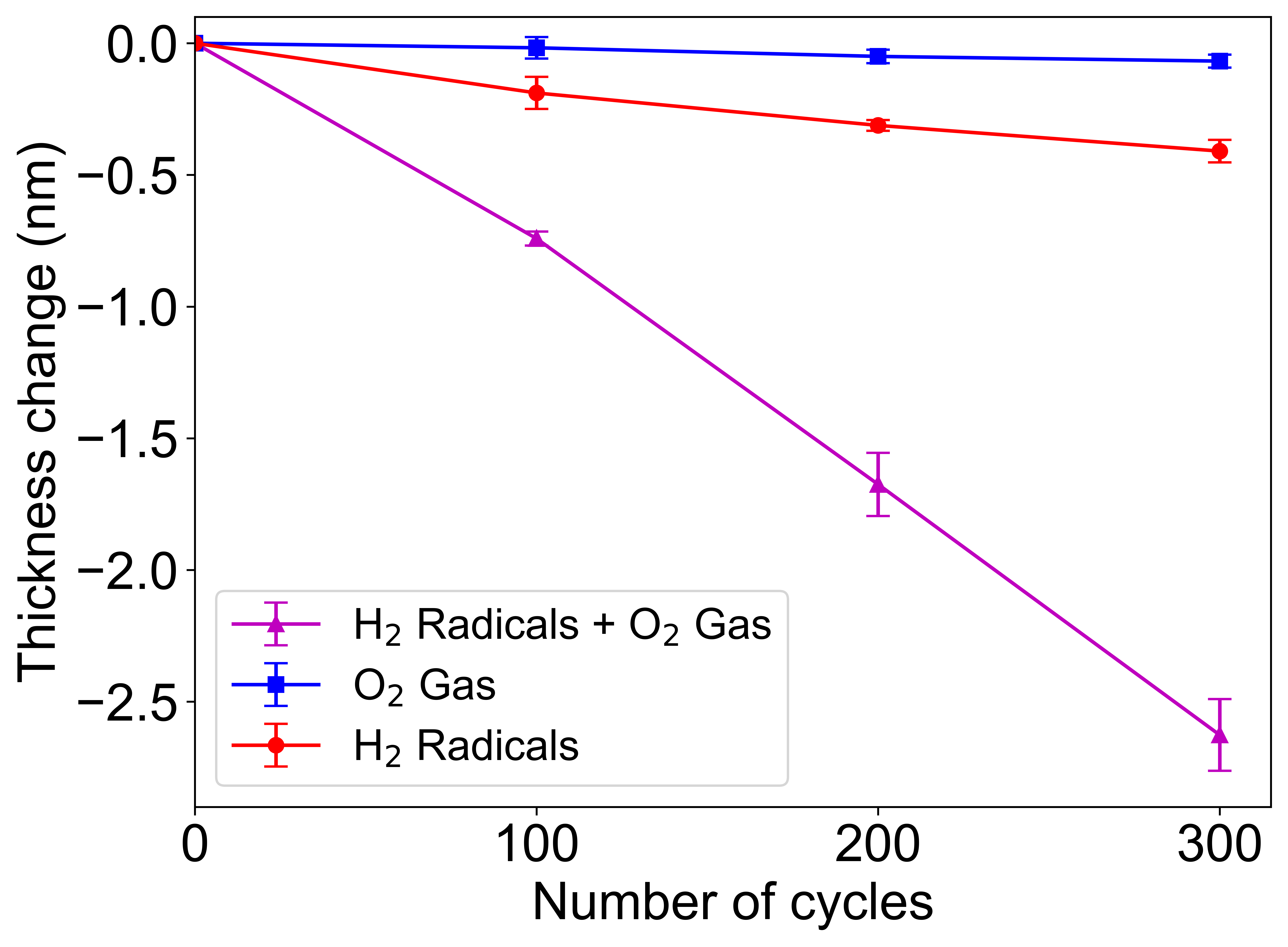}    }
    \caption{EPC versus cycle number for 45 seconds of O$_2$ gas per cycle (blue squares), 60 seconds of H$_2$ plasma (red circles), and both 45 seconds of O$_2$ gas and 60 seconds of H$_2$ radicals per cycle (purple triangles). All processes were carried out at $350$ \degreecelsius.   
    }
    \label{fig:Half-Cycle}
\end{figure}

We begin by comparing the change in sample thickness with respect to cycle number for oxygen gas, hydrogen radicals, and both half steps as shown in \Cref{fig:Half-Cycle}. We observe that the exposure of InGaAs to just oxygen gas has a negligible etch rate of 0.002 \AA/cycle. Exposure to only hydrogen radicals has a slightly higher EPC of 0.014 \AA/cycle. This EPC value can likely be attributed to the use of ex situ ellipsometry to measure the sample thickness, meaning that the sample is exposed to ambient air for some time prior to the measurement. As a result, some thickness of native oxide is formed which is then  etched by the next hydrogen radical exposure. This etched thickness is included in the EPC for hydrogen radicals. It is expected that the EPC excluding this effect is negligible. 

Combining the two half cycles, we observe an EPC of 0.095 \AA/cycle. The EPC was obtained by dividing the total thickness change after 300 cycles by the number of cycles. We note that to minimize process times, the sample was reused to obtain the thickness change after 100, 200, and 300 cycles, rather than using a new sample for each data point.  This EPC is similar to the previously reported EPC of thermal ALE of InGaAs using HF and DMAC for which an etch rate of 0.10 \AA/cycle at 250 \degreecelsius~was reported \cite{doi:10.1021/acs.nanolett.9b01525}.


\begingroup
\setlength{\jot}{2pt} 

The chemistry for this etching has been extensively studied in the context of ultrahigh vacuum surface cleaning of III-V semiconductors using hydrogen radicals. \cite{10.1116/1.1488949,10.1063/1.2919047} The generally accepted reactions are:


\begin{gather}
\mathrm{As_2O_x + 2xH \rightarrow xH_2O \uparrow + As_2\ (or\ \tfrac{1}{2}As_4) \uparrow} \\
\mathrm{Ga_2O_3 + 4H \rightarrow Ga_2O \uparrow + 2H_2O \uparrow} \\
\mathrm{In_2O_3 + 4H \rightarrow In_2O \uparrow + 2H_2O \uparrow}
\end{gather}

\endgroup

where the uparrow indicates volatile species under vacuum. These reactions are all known to proceed with volatile products at temperatures above 200 \degreecelsius \cite{10.1116/1.1488949,10.1063/1.2919047, 10.1063/1.1403684, hollowayhandbook}.

\begin{figure}
    \centering
    {\includegraphics[width=\textwidth]{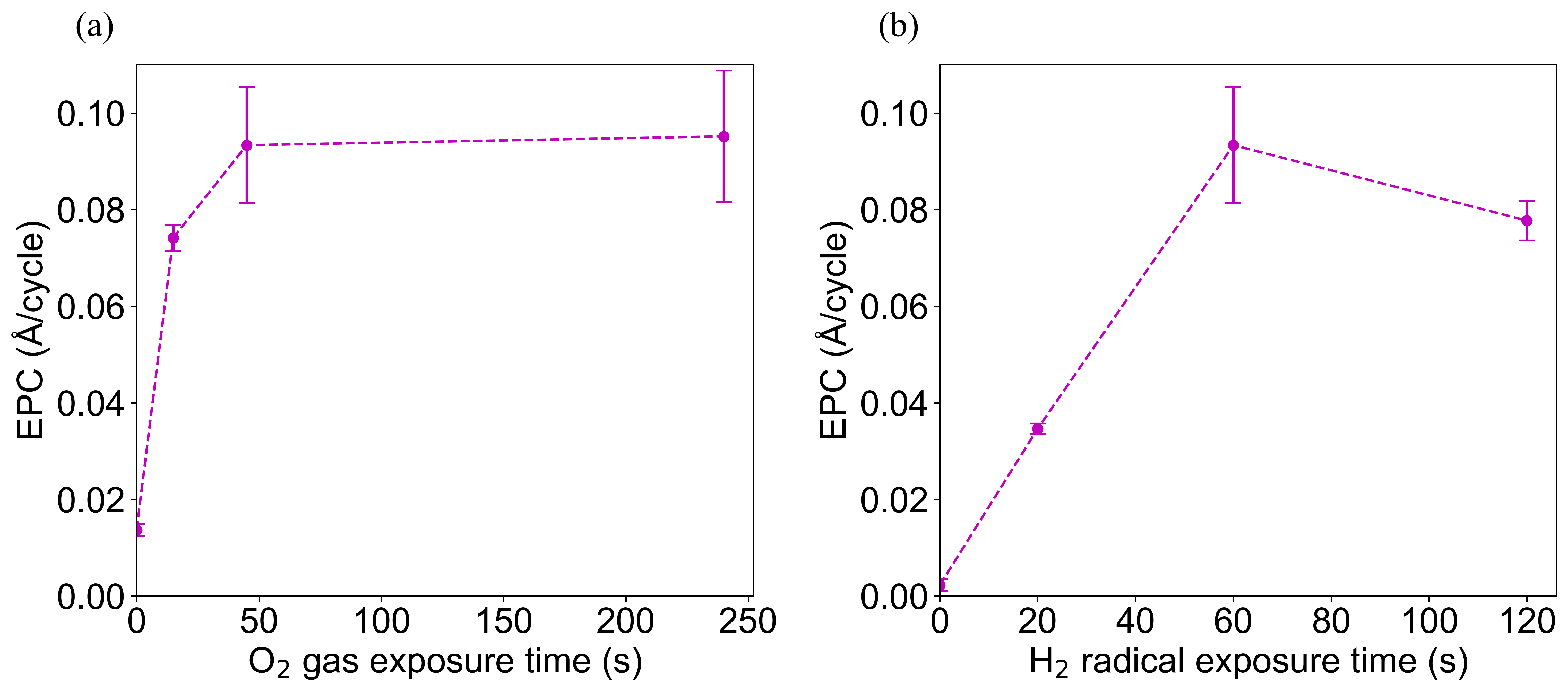}
    \phantomsubcaption\label{fig:o2-saturation}
    \phantomsubcaption\label{fig:h2-saturation}
    }
    \caption{(a) O$_2$ gas saturation curve obtained for a fixed H$_2$ plasma exposure of 60 seconds. (b) H$_2$ Plasma exposure time obtained for a fixed O$_2$ gas exposure of 45 seconds. The chamber temperature, pressure, pump and purge times were held constant for these plots.   
    }
    \label{fig:Saturation-Curve}
\end{figure}

To gain further insight into the process, we measured the saturation curves for both the hydrogen radical and oxygen gas half cycles. \Cref{fig:o2-saturation} shows the O$_2$ gas saturation curve. It is observed that the EPC increases with increasing O$_2$ gas exposure time until an exposure of 45 seconds is reached, after which the EPC plateaus. It is concluded that the oxygen half cycle has reached saturation at a 45 second per cycle exposure. Next, \Cref{fig:h2-saturation} shows the H$_2$ plasma saturation curve. This plot likewise shows that increasing the hydrogen plasma exposure time increases the EPC until 60 seconds after which it appears that the etch rate decreases slightly, although the EPC plateaus within uncertainty. This mild decrease could be due to  H ion damage as the ion flux is non-negligible for sufficiently long exposures, despite the use of the roof structure. However, considering prior studies of hydrogen radicals with native oxides of InGaAs, saturation of the hydrogen radical step is expected. We take the saturated ALE conditions to be 45 seconds of O$_2$ and 60 seconds of H$_2$ radicals for a process temperature of 350 \degreecelsius.


The temperature dependence of the EPC was surveyed by running 200 cycles with saturated parameters at $300$ \degreecelsius, $350$ \degreecelsius, and $400$ \degreecelsius. The EPC is found to increase by around 45\% at a temperature of 400 \degreecelsius, likely due to the increase in solid-state diffusivity of the oxygen gas molecules. The etch rate appeared negligible at 300 \degreecelsius~using the saturated process conditions at 350 \degreecelsius. A prior study has shown InGaAs oxides  may be removed as low as 192 \degreecelsius~with an exposure time of 10 minutes \cite{10.1063/1.345757}. Therefore, we expect that etching could be possible at lower temperatures with longer cycles times or higher hydrogen radical density. This topic will be investigated in a future study.

\begin{figure}
    \centering
    {\includegraphics[scale=.435]{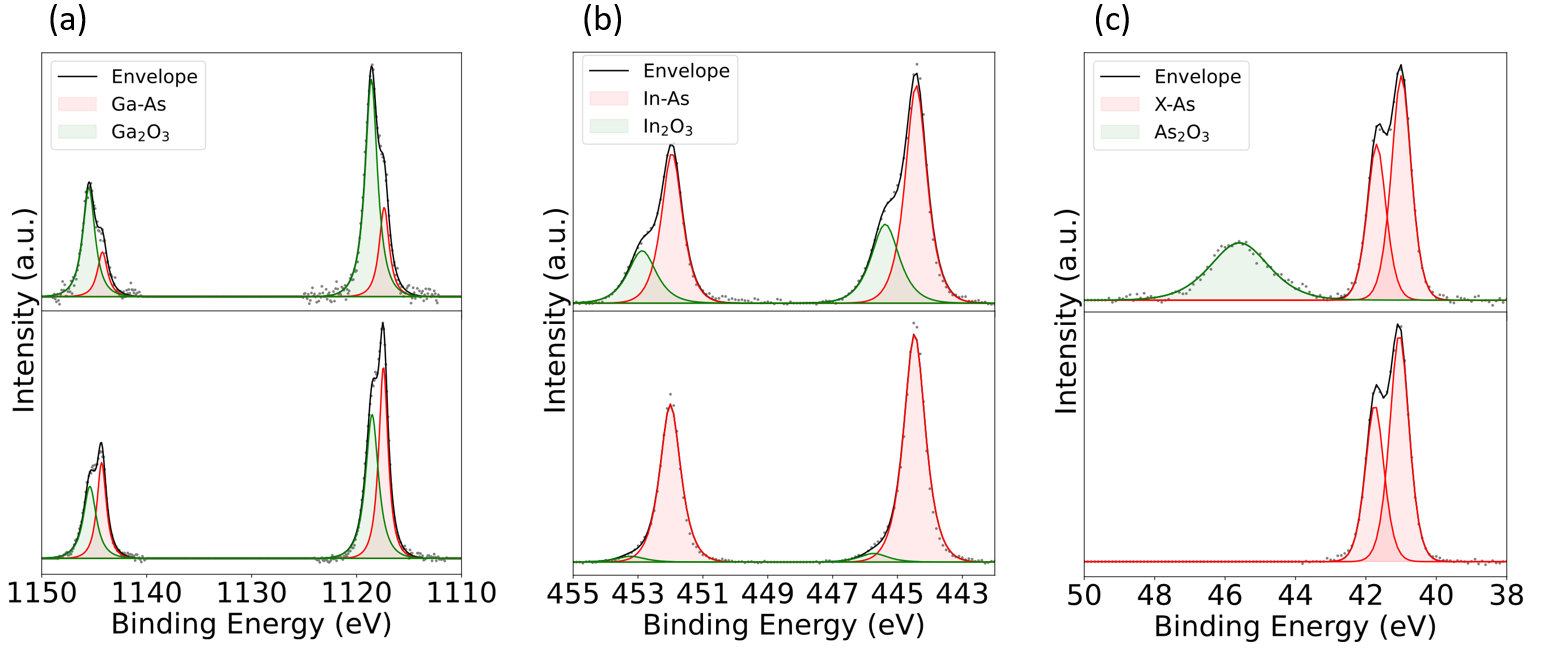}
    \phantomsubcaption\label{fig:Ga-xps}
    \phantomsubcaption\label{fig:In-xps}
    \phantomsubcaption\label{fig:As-xps}
    }
    \caption{Surface XPS spectra for (a) Ga2p, (b) In3d, and (c) As3d peaks. The spectra are shown for the original sample (top) and the etched sample (bottom). The measured spectra (grey dots) and fit spectra envelope (black line) are given in arbitrary units (a.u.) versus their binding energy (eV).       
    }
    \label{fig:Surface-XPS}
\end{figure}

Next, we analyze the surface chemical composition of the InGaAs samples before and after ALE using ex-situ XPS. \Cref{fig:Surface-XPS} shows the core level spectra for In3d, As3d, and Ga2p before and after 200 cycles of ALE. It is observed that the In and Ga core level spectra have 2 doublet peaks. There are two peaks due to the spin orbital splitting of these levels, 3d5/2 and 3d3/2 for In and 2p1/2, 2p3/2 for Ga, and they are doublets due to the fact these metals have two distinct chemical bonding environments: one metallic and one involving bonding with oxygen. The higher energy peak in the doublet corresponds to the oxides. Whereas, the doublet in the As spectra is due to spin orbital splitting (3d5/2 and 3d3/2) and the broad peak higher in energy corresponds to its oxidation states.

Qualitatively, it is seen that there are relatively fewer surface oxides after ALE despite the samples being exposed to ambient air prior to XPS. The proportion of indium oxide on the surface has also decreased relative to gallium oxide, and there is no detectable As oxide on the surface after ALE. It is difficult to determine if this difference is simply from the oxidation of the sample after leaving the etching chamber or if the  process conditions result in preferential etching which could potentially terminate etching over a large cycle number. This topic will be investigated in a future study.



\begin{figure}
    \centering
    {\includegraphics[width=\textwidth]{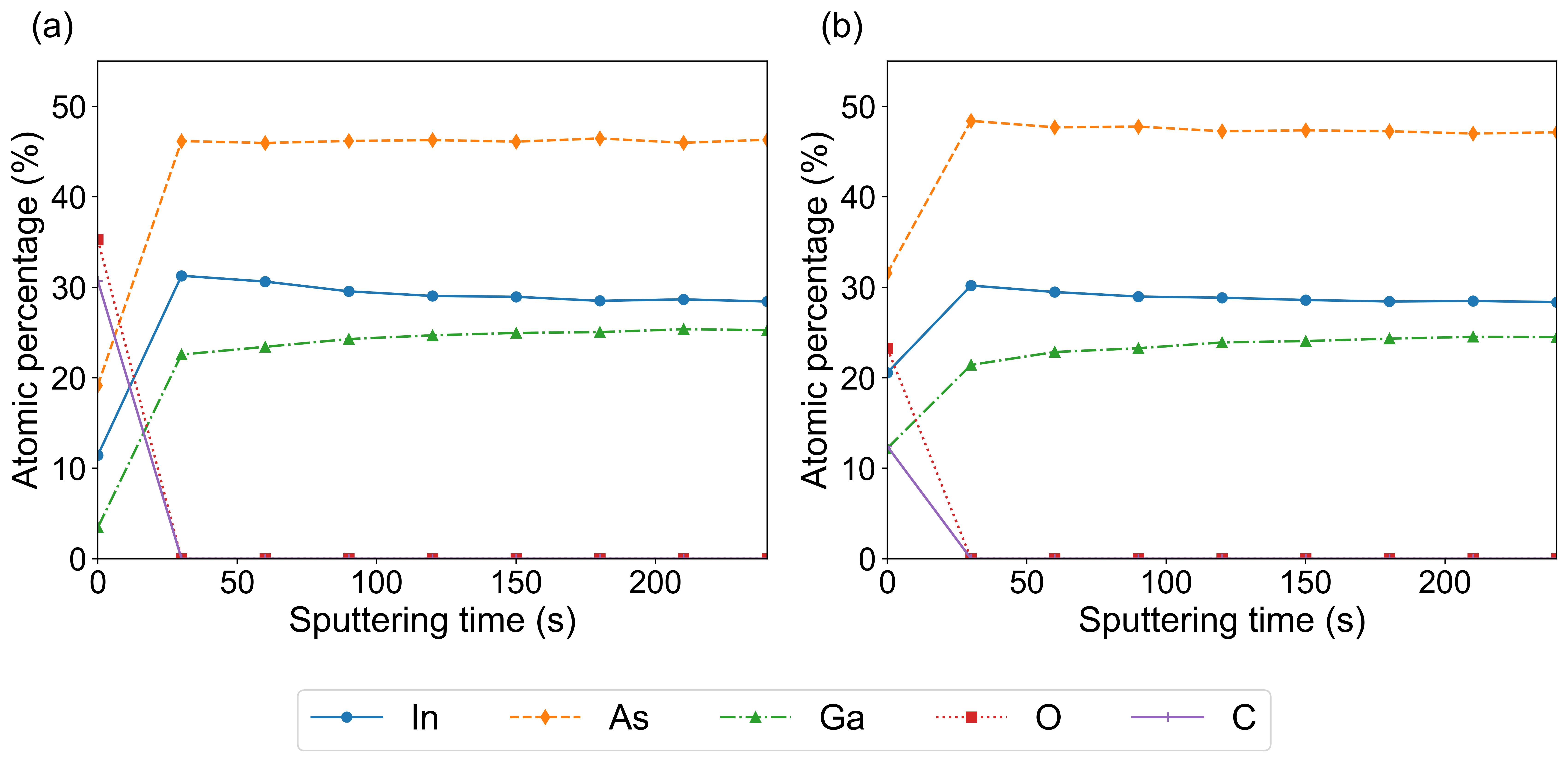}
    \phantomsubcaption\label{fig:depth-profile-pristine}
    \phantomsubcaption\label{fig:depth-profile-ale}
    }
    \caption{Atomic concentration of In, As, Ga, O, and C as a function of Ar milling time (s) for (a) original sample with native oxide, (b) sample with 200 cycles of ALE.     
    }
    \label{fig:Depth-Profile}
\end{figure}

\begin{table}
\caption{\label{tab:xps_table} Surface atomic composition from XPS.}
\begin{ruledtabular}
\begin{tabular}{cccccc}
  & In (\%) & As (\%) & Ga (\%) & O (\%) & C (\%) \\
\hline
Original   & 11.43 $\pm$ 0.07 & 19.16 $\pm$ 0.27 & 3.45 $\pm$ 0.06 & 35.26 $\pm$ 1.34 & 30.70 $\pm$ 2.41\\
ALE   & 20.57 $\pm$ 0.08 & 31.56 $\pm$ 0.35 & 12.21 $\pm$ 0.22 & 23.26 $\pm$ 2.85 & 12.41 $\pm$ 1.22
\end{tabular}
\end{ruledtabular}
\end{table}

To assess the degree to which the processing is confined to the surface region,  XPS depth profiling was performed. The atomic concentration of O, C, In, Ga, and As with respect to milling time for both the original sample and a sample etched with 200 cycles of ALE are shown in \Cref{fig:Depth-Profile}. The bulk atomic composition after ALE was observed to be within 97\% of the values of the original sample. Furthermore, the bulk ratio of In:Ga was around 1.13. The surface composition of both the original and the ALE sample is shown in \Cref{tab:xps_table}. Here, it is observed that after ALE the surface atomic concentration of oxygen decreases by 34.0\%. It was also observed that the ratio of In:Ga was halved after ALE. However, the In:Ga ratio after ALE, 1.7, is still above the bulk ratio. It should be noted that the original sample surface was In rich. It is unclear if this apparent decrease in In from ALE will continue past the bulk stoichiometry for higher cycle number. 

\begin{figure}
    \centering
    {\includegraphics[width=0.95\textwidth]{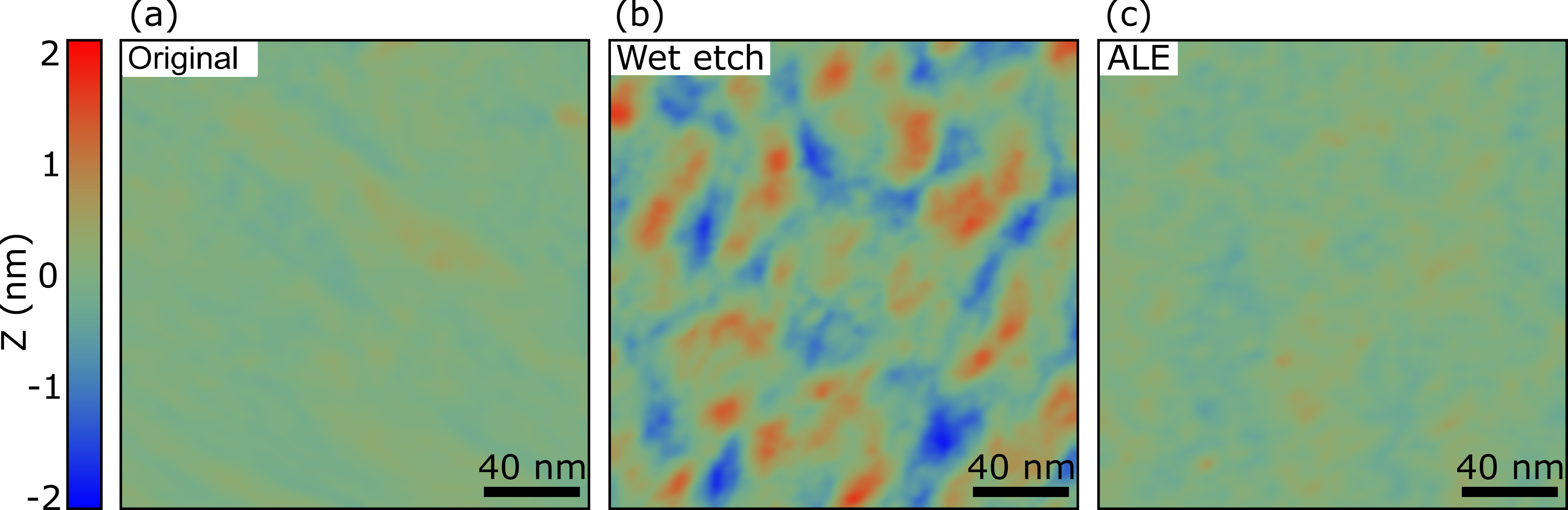}
    \phantomsubcaption\label{fig:pristine-roughness}
    \phantomsubcaption\label{fig:wet-etch-roughness}
    \phantomsubcaption\label{fig:ale-roughness}
    }
    \caption{Surface height map of (a) original, (b) wet etched, and (c) ALE InGaAs. The RMS roughnesses are (a) $R_q = 0.11$ nm, (b) 0.51 nm, and (c) 0.13 nm, respectively.          
    }
    \label{fig:AFM-Roughness}
\end{figure}

We used AFM to characterize the surface roughness of a original InGaAs sample with its native oxide, a sample subjected to 200 cycles of ALE (etch depth of 2 nm), and a sample etched for $\sim$ 5 s (etch depth of $\sim 6-10$ nm) with a citric acid/H$_2$O$_2$ wet etch. \Cref{fig:AFM-Roughness} shows a height map of the surface of these samples. The original sample (\Cref{fig:pristine-roughness}) has a $R_q=0.11$ nm, as expected for an epitaxially grown film. The wet etched sample (\Cref{fig:wet-etch-roughness}) is visibly rougher with an $R_q=0.51$ nm. The ALE sample (\Cref{fig:ale-roughness}) has an $R_q=0.13$ which is indistinguishable from that of the MBE grown sample. 

\begin{figure}
    \centering
        {\includegraphics[scale=.55]{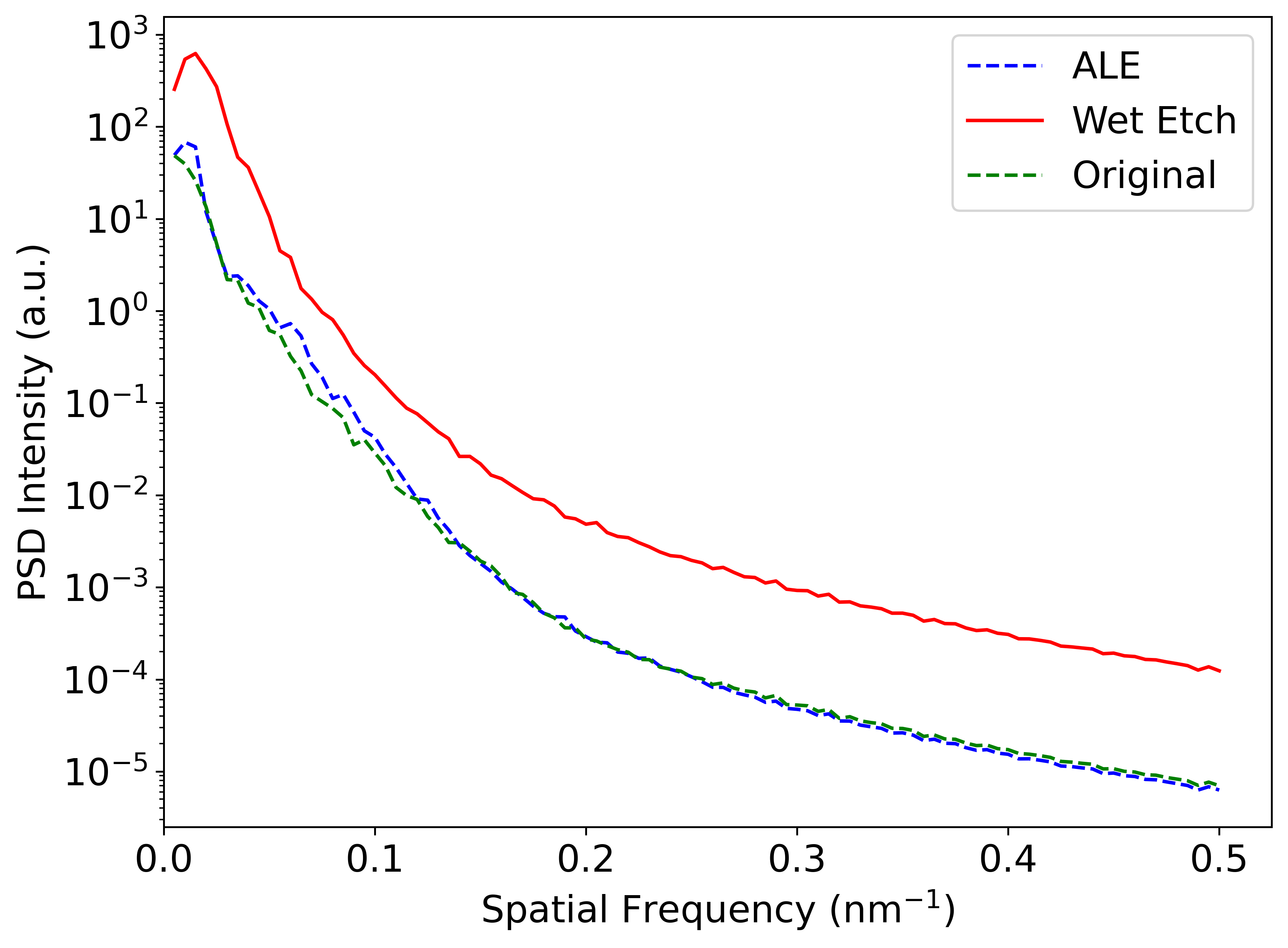}}
    \caption{PSD obtained from AFM scans of original (green dashed), wet etched (red dashed), and ALE InGaAs (blue dashed).         
    }
    \label{fig:Power Spectral-Density}
\end{figure}

Power spectral density (PSD) is another way to quantify surface roughness over characteristic length scales as quantified by the spatial frequency.  The PSD of the AFM scans is shown in \Cref{fig:Power Spectral-Density}. It is observed that the wet-etched sample has a higher PSD over all spatial frequencies compared to the original sample. On the other hand, the ALE sample exhibits a PSD which nearly coincides with that of the original sample. This result further confirms that 200 cycles of ALE does not increase the surface roughness. 

\section{Discussion}
Our ALE process has potential applications for III-V compound semiconductor devices by enabling precise etch control on the Angstrom scale with wafer-scale uniformity. Unlike continuous wet or dry etching processes which are subject to local variations in etch rate over a wafer, ALE is inherently scalable to arbitrary wafer sizes owing to the use of self-limiting surface chemical reactions. Our process may find application as gate recess etching for InP HEMTs where sub-nanometer precision in the etch depth over a 3'' or larger InP wafer is desired for optimal yield. The low surface damage of ALE compared to traditional wet or dry etches may facilitate more aggressive device scaling, leading to improved high frequency performance. A lower electrical trap density on the ALE-treated surface is also expected, which could lead to improved gain stability. 

Future work should employ a downstream plasma or thermal source of hydrogen radicals to eliminate ion damage while also providing a higher radical density. Additional study is needed to assess whether the process also etches the InAlAs barrier film. Finally, surface passivation schemes for the exposed InAlAs need to be developed. ALE offers the possibility for in-situ passivation of this surface in the clean environment of high vacuum, potentially via formation of a crystalline oxide followed by atomic layer deposition, \cite{10.1063/1.4979202} or other schemes.



\section{Conclusion}
We have reported an ALE process for InGaAs consisting of sequential hydrogen radical and oxygen gas exposures. We observe an etch rate of 0.095 \AA/cycle and that both half-cycles exhibit self-limiting behavior. The process preserves the atomic-scale smoothness of the original epitaxial wafer after 200 cycles of ALE. The process has potential for applications in fabrication of InP HEMTs with improved high frequency performance and gain stability.


\section{Acknowledgements}
The authors thank Andrew Carter for useful discussions. This prototype (or technology) was partially supported by the Microelectronics Commons Program, a DoD initiative, under award number N00164-23-9-G056, and the Rothenberg Innovation Initiative (RI\textsuperscript{2}) at the California Institute of Technology. This material is based upon work supported by the National Science Foundation under Grant No.\ 191122. We gratefully acknowledge the critical support and infrastructure provided for this work by The Kavli Nanoscience Institute at the California Institute of Technology. Research was in part carried out at the Molecular Materials Research Center in the Beckman Institute of the California Institute of Technology. This work was performed, in part, at the Center for Integrated Nanotechnologies, an Office of Science User Facility operated for the U.S. Department of Energy (DOE) Office of Science. Sandia National Laboratories is a multimission laboratory managed and operated by National Technology \& Engineering Solutions of Sandia, LLC, a wholly owned subsidiary of Honeywell International, Inc., for the U.S. DOE’s National Nuclear Security Administration under contract DE-NA-0003525. The views expressed in the article do not necessarily represent the views of the U.S. DOE or the United States Government

\section{Data availability statement}

The data that support the findings of this study are available from the corresponding author upon reasonable request.

\section{Conflict of interest}

The authors have no conflicts to disclose.

\newpage
\clearpage

\bibliographystyle{is-unsrt}
\bibliography{ref,AA_references}

\end{document}